\documentclass[a4paper,lnbip]{svmultln}
\usepackage{graphicx}
\usepackage{float}
\usepackage{threeparttable}
\usepackage{multirow}
\usepackage{wrapfig}

\graphicspath{{img/}}

\begin{document}
\mainmatter

\title{Modeling Styles in Business Process Modeling\thanks{The final publication
is available at Springer via http://dx.doi.org/10.1007/978-3-642-31072-0\_11}}
\titlerunning{Modeling Styles in Business Process Modeling}
\author{Jakob Pinggera\inst{1} \and Pnina Soffer\inst{2} \and Stefan
Zugal\inst{1} \and Barbara Weber\inst{1} \and Matthias Weidlich\inst{3} \and Dirk
Fahland\inst{4} \and Hajo A. Reijers\inst{4} \and Jan Mendling\inst{5}}
\authorrunning{Jakob Pinggera et al.}

\institute{University of Innsbruck, Austria\\
\email{{jakob.pinggera, stefan.zugal, barbara.weber}@uibk.ac.at} \and
	University of Haifa, Israel\\
\email{spnina@is.haifa.ac.il} \and
	Eindhoven University of Technology, The Netherlands\\
\email{d.fahland|h.a.reijers@tue.nl} \and
	Technion - Israel Institute of Technology\\
\email{weidlich@tx.technion.ac.il} \and
	Vienna University of Economics and Business\\
\email{jan.mendling@wu.ac.at}
} \maketitle

\begin{abstract}
Research on quality issues of business process models has recently begun to
explore the process of creating process models. As a consequence, the
question arises whether different ways of creating process models exist. In
this vein, we observed 115 students engaged in the act of modeling, recording
all their interactions with the modeling environment using a specialized
tool. The recordings of process modeling were subsequently clustered. Results
presented in this paper suggest the existence of three distinct modeling
styles, exhibiting significantly different characteristics. We believe that
this finding constitutes another building block toward a more comprehensive
understanding of the process of process modeling that will ultimately enable
us to support modelers in creating better business process models.
\end{abstract}

\vspace{-0.8cm}
\section{Introduction}
Considering the heavy usage of business process modeling in all types of business
contexts, it is important to acknowledge both the relevance of process models and
their associated quality issues. However, actual process models display a
wide range of problems~\cite{mendling-lnbip}. Following the SEQUAL
framework~\cite{Lin+94}, quality dimensions of models include syntactic,
semantic, and pragmatic quality. Syntactic and semantic quality relate to model
construction, and address the correct use of the modeling language and the extent
to which the model truthfully represents the real world behavior it should
depict, respectively. Pragmatic quality addresses the extent to which a model
supports its usage for purposes such as understanding behavior or developing
process aware systems. Considering process models whose purpose is to develop an
understanding of real world behavior, pragmatic quality is typically related to
the understandability of the model~\cite{krogstie06}. Clearly, an in-depth
understanding of the factors influencing the various quality dimensions of
process models is in demand.

Most research in this area puts a strong emphasis on the product or outcome of
the process modeling act (e.g.,~\cite{van2000verification,gruhn2006complexity}).
For this category of research, the resulting model is the object of analysis.
Many other works---instead of dealing with the quality of individual
models---focus on the characteristics of modeling languages
(e.g.,~\cite{siau2008evaluation,DBLP:journals/tse/Moody09}). Recently, research
has begun to explore another dimension presumably affecting the quality of
business process models by incorporating the process of creating a process model
into their investigations (e.g.,~\cite{Sof+11,PZW+11}). In particular, the focus
has been put on the formalization phase in which a process modeler is facing the
challenge of constructing a syntactically correct model reflecting a given domain
description (cf.~\cite{hoppenbrouwers05er}). Our research can be attributed to
the latter stream of research.

This paper contributes to our understanding of the process of process modeling
(PPM) by investigating whether different ways of process modeling can be
identified, i.e., can we observe different modeling styles when modelers create
process models? Knowledge about different modeling styles will support the
creation of customized process modeling environments, supporting modelers in
creating high quality models. Similarly, a more comprehensive understanding of
the PPM can be exploited for teaching students in how to create process models of
high quality. We conducted a modeling session with 115 students, recording all
their interactions with the modeling environment using a specialized tool. To
identify different modeling styles the collected PPM instances were automatically
clustered suggesting the existence of three different modeling styles. The
modeling styles were subsequently analyzed using a series of measures for
quantifying the PPM to validate differences between the three groups.

The paper is structured as follows. Section~\ref{sec:background} presents
backgrounds on the PPM and introduces measures for
quantifying this process. Section~\ref{sec:clustering} describes data collection
and cluster analysis. Section~\ref{sec:results} presents the results,
followed by their discussion in Section~\ref{sec:interpretation}. The paper is
concluded with a discussion of related work in Section~\ref{sec:relatedWork} and
a brief summary in Section~\ref{sec:summary}.

\vspace{-0.5cm}
\section{Backgrounds}\label{sec:background}
This section provides background information on the PPM
and explains how this process can be captured and quantified using a series of
measures.

\subsection{The Process of Process Modeling}

During the formalization phase process modelers are working on creating a
syntactically correct process model reflecting a given domain description by
interacting with the process modeling tool \cite{hoppenbrouwers05er}. This
modeling process can be described as an iterative and highly flexible
process~\cite{Cra+00,Morr67}, dependent on the individual modeler and the
modeling task at hand~\cite{Will95}. At an operational level, the modeler's
interactions with the tool would typically consist of a cycle of the three
successive phases of (1) comprehension (i.e.,\ the modeler forms a mental
model of domain behavior), (2) modeling (i.e.,\ the modeler maps the mental
model to modeling constructs), and (3) reconciliation (i.e.,\ the modeler
reorganizes the process model)~\cite{PZW+11,Sof+11}.

\noindent\textbf{Comprehension.} Research on human cognition and problem
solving has shed light on comprehension. According to~\cite{NeSi72}, when facing
a task, the problem solver first formulates a mental representation of the
problem, and then uses it for reasoning about the solution and which methods
to apply for solving the problem. In process modeling, the task is to create
a model which represents the behavior of a domain. The process of forming
mental models and applying methods for achieving the task is not done in one
step applied to the entire problem. Rather, due to the limited capacity of
working memory, the problem is broken down to pieces that are addressed
sequentially, chunk by chunk~\cite{Sof+11,PZW+11}.

% We turn to research in the area of human
% cognition and problem solving for guidance in gaining better understanding of the
% cognitive processes involved in model creation. According to~\cite{NeSi72}, when
% facing a task, the problem solver first formulates a mental representation of the
% problem, also termed "the problem space", and then uses it for reasoning about
% the solution. The cognitive fit theory~\cite{Kha+06,Vesse91} adopts on this view,
% stressing that matching information types along this process support high
% performance in the problem solving task. In process modeling, the task is to
% create a model which represents the behavior of a domain. Two characteristics of
% problem solving, indicated by~\cite{NeSi72}, are of particular interest. First,
% the shape of the mental model is affected by the characteristics of the task and
% the methods for achieving it. Hence, the concepts available to the modeler for
% reasoning about the domain may affect the mental modeling process even before an
% actual mapping to constructs is performed. Second, the process of forming mental
% models and applying methods for achieving the task is not done in one step
% applied to the entire problem. Rather, due to the limited capacity of working
% memory, the problem is broken down to pieces that are addressed sequentially,
% chunk by chunk.

\noindent\textbf{Modeling.} The modeler uses the problem and solution
developed in working memory during the previous comprehension phase to
materialize the solution in a process model (by creating or changing
it)~\cite{Sof+11,PZW+11}. The modeler's utilization of working memory
influences the number of modeling steps executed during the modeling phase
before forcing the modeler to revisit the problem for acquiring more
information~\cite{PZW+11}.

\noindent\textbf{Reconciliation.} After modeling, modelers typically
reorganize the process model (e.g.,\ renaming of activities) and utilize the
process model's \emph{secondary notation} (e.g.,\ notation of layout,
typographic cues) to enhance the process model's
understandability~\cite{Petr95,Men+07}. However, the number of reconciliation
phases in the PPM is influenced by a modeler's ability of placing elements
correctly when creating them, alleviating the need for additional
layouting~\cite{PZW+11}.

\subsection{Capturing Events of the Process of Process Modeling}
\label{sect:operations}

To investigate the PPM, actions taken during modeling have to be recorded and
mapped to the phases described above. When modeling in a process modeling
environment, process modeling consists of adding nodes and edges to the
process model, naming or renaming activities, and adding conditions to edges.
In addition to these interactions a modeler can influence the process model's
secondary notation, e.g.,\ by laying out the process model using move
operations for nodes or by utilizing bendpoints to influence the routing of
edges, see~\cite{PZW+11} for details.

To capture modeling activities, and for obtaining a closer look on how
process models are created in a systematic manner, we instrumented a basic
process modeling editor to record each user's interactions together with the
corresponding time stamp in an event log, describing the creation of the
process model step by step. Editor and event recording are available within
Cheetah Experimental Platform (CEP)~\cite{PiZW10}.

\subsection{Quantifying the Process of Process Modeling}\label{sec:measures}

A log of modeling events allows quantitative analysis of a PPM. Based on the
conceptual background, comprehension (C), modeling (M), and reconciliation
(R) phases can be identified by grouping events into respective phases
(see~\cite{PZW+11} for details). Then, a PPM can be divided into
\emph{modeling iterations}~\cite{PZW+11}. One iteration is assumed to
comprise a comprehension (C), modeling (M), and reconciliation (R) phase in
this respective order. The iterations of a modeling process are identified by
aligning its phases to the CMR-pattern. If a certain phase of this pattern is
not present in the modeling process, the respective phase is skipped for the
observed iteration and the process is considered to continue with the next
phase of the pattern. In the following we present five measures quantifying
the process of process modeling.

\noindent \textbf{Number of Iterations.} This measure counts the modeling
iterations per PPM reflecting how often a modeler had to interrupt
modeling for comprehension or reconciliation.

\noindent \textbf{Share of Comprehension.} When comprehending, a mental model
of the problem and a corresponding solution is developed which is then
formalized in modeling phases. Differences in the amount of time spent on
comprehension can be expected to characterize modeling styles and to impact
on the modeling result. We quantify this aspect as the ratio of the average
length of a comprehension phase in a process to the average length of an
iteration. The initial comprehension phase is neglected as it is typically
subject to various influences unrelated to problem solving (e.g., the modeler
did not start immediately).

\noindent \textbf{Iteration Chunk Size.} Modelers can be assumed to conduct
modeling in chunks of different sizes. We quantified \emph{chunk size} as the
average number of create and delete operations executed in one iteration. This
measure reflects the ability to model large parts of a model without the need to
comprehend or reconcile.

\noindent \textbf{Reconciliation Breaks.} A steady process of modeling is
assumed to be a sequence of iterations following the CMR-pattern. Reconciliation
can sometimes be skipped if the modeler can place all model elements directly
at the right spot clearly alleviating the need for reconciliation. However,
some processes may even show iterations of CR-patterns, i.e., an iteration
without a modeling phase, where a modeler interrupts the common flow of
modeling for additional reconciliation. We quantified this aspect by the
relative share of iterations that comprise unexpected reconciliation (without
modeling) out of all iterations.

\noindent \textbf{Delete Iterations.} From time to time, modelers are
required to remove content from the process model. This might happen when
modelers identify errors in the model, which are subsequently resolved by
removing some of the modeling constructs and implementing the desired
functionality. This measure describes the number of iterations of the PPM
containing delete operations relative to the total number of iterations of
the PPM.

\vspace{-0.2cm}
\section{Clustering}\label{sec:clustering}
To be able to make generalizations, we have used cluster analysis to a set of PPM
instances. Cluster analysis allows us to identify groups of modelers
exhibiting similar modeling styles. This section describes the modeling
session, data pre-processing and cluster analysis.

\subsection{Data Collection}
The modeling session was designed to collect PPM instances of students
creating a formal process model in BPMN from an informal description.
The object that was to be modeled is a process describing the activities a pilot
has to execute prior to taking off an aircraft\footnote[1]{Material download:
http://pinggera.info/experiment/ModelingStyles}.

To mitigate the risk that the PPM instances were impacted by
complicated tools or notations~\cite{Cra+00}, we decided to use a subset of BPMN
for our experiment. In this way, modelers were confronted with a minimal number
of distractions, but the essence of how process models are created could still be
captured. A pre-test was conducted at the University of Innsbruck to ensure the
usability of the tool and the understandability of the task description. This led
to further improvements of CEP and minor updates to the task description.

The modeling sessions were conducted in November 2010 with students of a
graduate course on Business Process Management at Eindhoven University of
Technology and in January 2011 with students from Humboldt-Universit\"{a}t zu
Berlin following a similar course. The modeling session at each university
started with a demographic survey, followed by a modeling tool tutorial
explaining the basic features of CEP. After that, the actual modeling task was
presented in which the students had to model the above described ``Pre-Flight''
process. This was done by 102 students in Eindhoven and 13 students in Berlin. By
conducting the experiment during class and closely monitoring the students, we
mitigated the risk of falsely identifying comprehension phases due to external
distractions. No time restrictions were imposed on the students.

\subsection{PPM Profile for Clustering}

When trying to identify different types of PPM instances using clustering,
the question arises how to represent such a process to make clustering
possible. Based on our previous experience with the PPM we decided to focus
on four aspects. The adding of content, the removal of content,
reconciliation of the model and comprehension time, i.e.,\ the time when the
modeler does not work on the process model. To also reflect that modeling is
a time-dependent process, we do not just look at the total amount of modeling
actions and comprehension, but on their \emph{distribution} over time as
follows. We sampled every process into segments of $10$ seconds length. For
each segment, we compute its \emph{profile} $(a,d,r,c)$, i.e.,\ the numbers
$a$, $d$, and $r$ of add, delete, and reconciliation events, and the time $c$
spent on comprehension. The \emph{profile} of one PPM is then sequence
$(a_1,d_1,r_1,c_1)(a_2,d_2,r_2,c_2)\ldots$ of its segments' profiles. The
$a$, $d$, and $r$ are obtained per segment by classifying each event
according to \tablename~\ref{tab:userInteractions}. Adding a condition to an
edge was considered being part of creating an edge. Comprehension time
$c$ was computed as follows. Group events to intervals: an interval is a
sequence of events where two consecutive events are $\leq 1$ second apart,
its duration is the time difference between its first and its last event
(intervals of 1 activity got a duration of 1 second). Then $c$ is the length
of the segment (10 secs) minus the duration of all intervals in the segment.
For example, if the modeler moved activity A after $3$ secs, activity B after
$3.5$ secs and activity C after $4.2$ secs the comprehension time in this
segment would be $10 - 1.2 = 8.8$ seconds. To give all PPM profiles equal
length, shorter profiles were extended with segments of no interaction to
reach the length of the longest PPM (required for clustering). 

%\vspace{-0.3cm}
\begin{table}[t]\small\sf
\begin{center}
\begin{tabular}{@{\extracolsep{.2em}}llll} 
\hline
\textbf{Interaction} & \textbf{Classification} & \textbf{Interaction} & \textbf{Classification} \\
\hline
CREATE NODE & Adding & RENAME ACTIVITY & Reconciliation \\
DELETE NODE & Deleting & UPDATE CONDITION & Reconciliation  \\
CREATE EDGE & Adding &  MOVE NODE & Reconciliation  \\
DELETE EDGE & Deleting & MOVE EDGE LABEL & Reconciliation\\
RECONNECT EDGE & Adding/Deleting & MODIFY EDGE BENDPOINT  & Reconciliation \\
\hline
\end{tabular}
  \caption{Classification of CEP's User Interactions}
  \label{tab:userInteractions}
\end{center}
\end{table}

\subsection{Clustering}

The PPM profiles were exported from CEP~\cite{PiZW10} and subsequently
clustered using Weka\footnote{http://www.cs.waikato.ac.nz/ml/weka}. The
KMeans algorithm, first proposed in~\cite{macq67}, utilizing an euclidean
distance measure was chosen for clustering as it constitutes a well known and
easy to use means for cluster analysis. As KMeans might converge in a local
minimum~\cite{HaEl02}, the obtained clustering has to be validated. If the
identified clusters exhibit significant differences with regard to the
measures described in Section~\ref{sec:background}, we conclude that
different modeling styles were identified. KMeans requires the number of
cluster to be known a priori. As this was not the case we gradually
increased the number of clusters starting from 2, resulting in only one major
cluster. Setting the number of expected clusters to 3 revealed two major
clusters and one cluster of 2 PPM instances. Most promising results were
achieved by setting the number of clusters to be generated to 4 and starting
with a seed of 10, returning 3 major clusters and one small cluster of 2 PPM
instances. We considered these 3 major clusters for further analysis;
increasing the number of expected clusters only generated additional small
clusters.

\vspace{-0.2cm}
\section{Results}\label{sec:results}
In this section we present results of the cluster analysis and validate the
difference among the clusters using the measures described in
Section~\ref{sec:background}.

\subsection{Three Clusters}

\begin{figure}[h!tb] 
\begin{center}
  \includegraphics[width=0.97\textwidth]{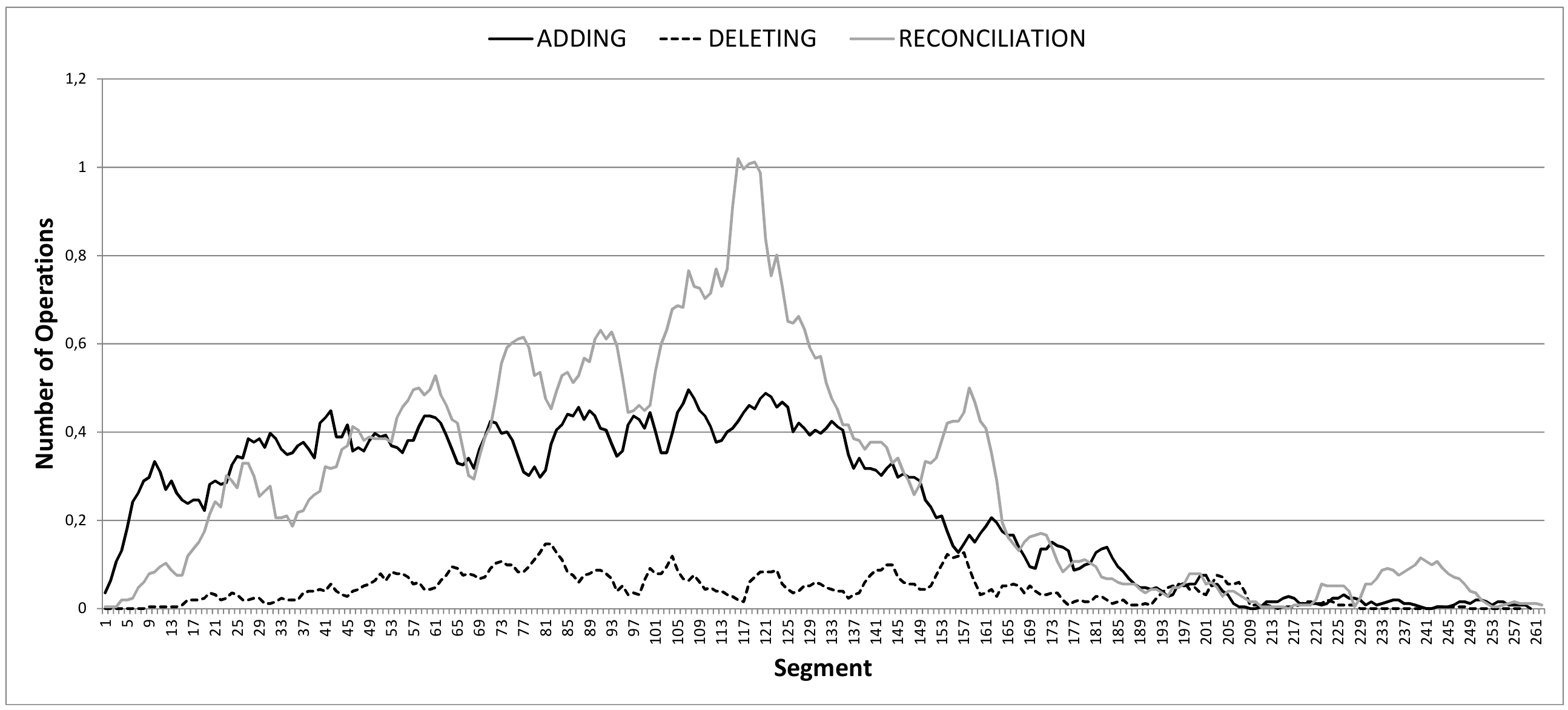}
  \caption{Cluster C1}
  \label{fig:cluster1}\par\vspace{\textfloatsep}
  %%%
  \includegraphics[width=0.97\textwidth]{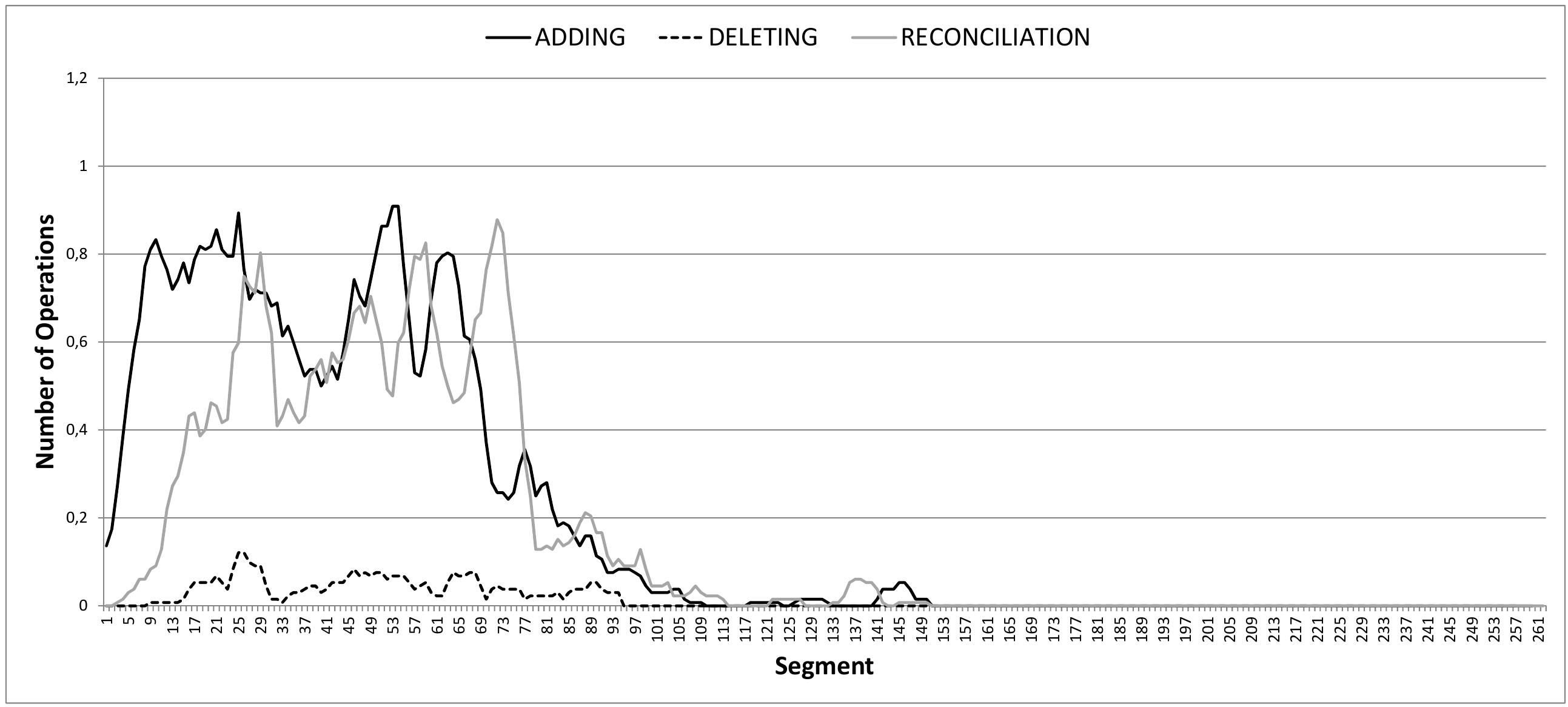}
  \caption{Cluster C2}
  \label{fig:cluster2}\par\vspace{\textfloatsep}
  %%%
  \includegraphics[width=0.97\textwidth]{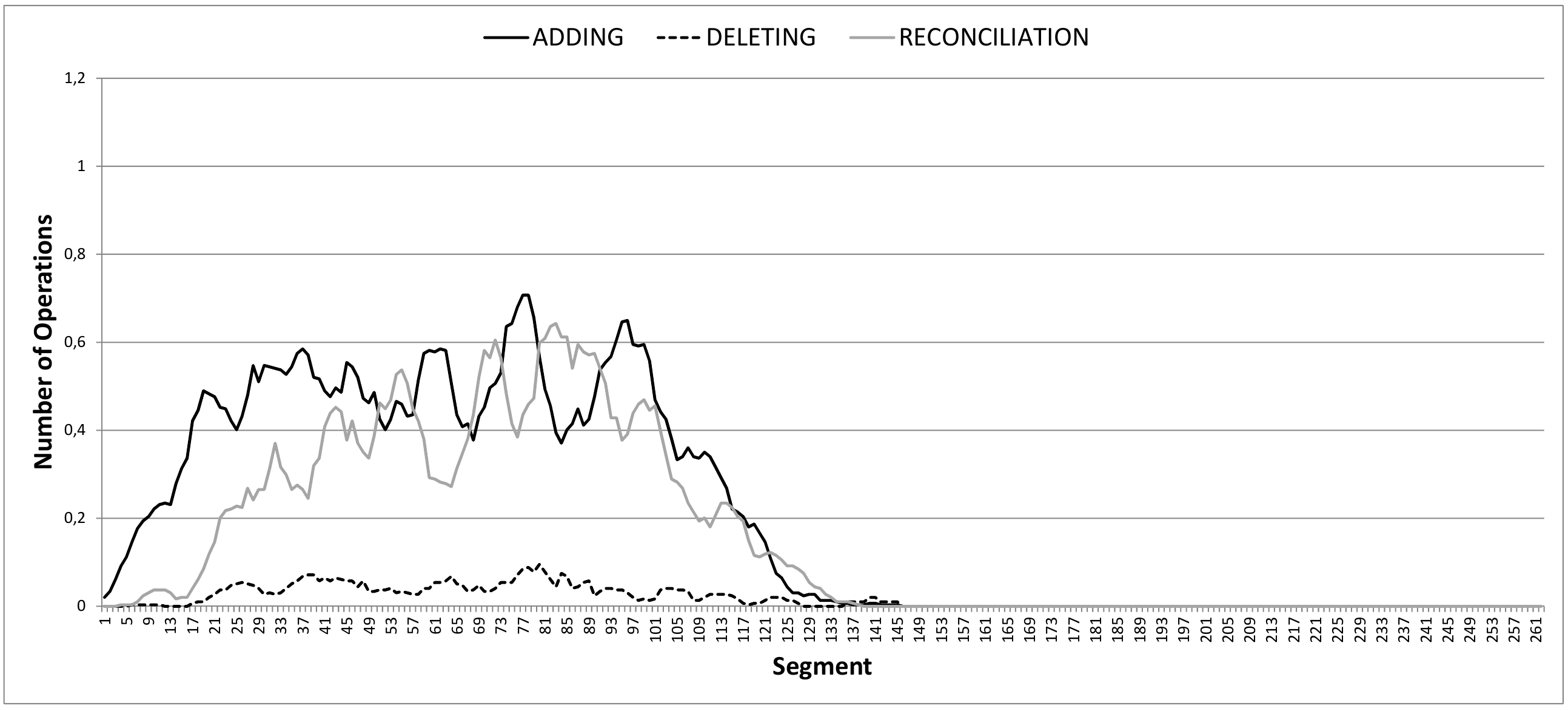}
  \caption{Cluster C3}
  \label{fig:cluster3}
\end{center}
\end{figure}

We identified three major clusters of 42, 22 and 49 instances, called C1, C2,
and C3 in the sequel. In order to visualize the obtained clusters we
calculated the average number of adding, deleting and reconciliation
operations per segment for each cluster. Additionally, we calculated the
moving average of six segments, i.e.,\ one minute, providing us with a
smoother representation of the modeling processes presented in
Figures~\ref{fig:cluster1}, \ref{fig:cluster2}, and \ref{fig:cluster3} for
C1, C2, and C3 respectively. The horizontal axis denotes the segments into
which the PPM instances were sampled. The vertical axis indicates the average
number of operations that were performed in this segment. For example, a
value of 0.8 for segment 9 (cf. \figurename~\ref{fig:cluster2}) indicates
that all modelers in this cluster averaged 0.8 adding operations within this
10 second segment.

\begin{table}[t]\small\sf
\centering
\begin{tabular}{@{\extracolsep{1em}}lrrr}
\hline
\textbf{Measure} & \textbf{C1} & \textbf{C2} & \textbf{C3} \\
\hline
Number of instances & 42 & 22 & 49 \\
Avg. no. of adding operations & 61.36 & 52.91 & 52.57  \\
Avg. no. of deleting operations & 10.81 & 3.91 &  4.55  \\
Avg. no. of reconciliation operations & 76.26 & 42.00 & 39.27 \\
Avg. no. operations & 148.43 & 98.82 & 96.39 \\
\hline
\end{tabular}
\caption{Statistics per cluster}
\label{tab:statistics}
\end{table}

C1 (cf. \figurename~\ref{fig:cluster1}) is characterized by long PPM
instances, as the first time the adding series reaches 0 is after about 205
segments. Additionally, the delete series indicates more delete operations
compared to the other clusters. Several fairly large spikes of reconciliation
activity can be observed, the most prominent one after about 117 segments.

C2, as illustrated in \figurename~\ref{fig:cluster2}, is characterized by a
fast start as a peak in adding activity is reached after 13 segments. In
general, the adding series is most of the time between 0.5 and 0.9
operations, higher compared to the other two clusters. The fast modeling
behavior results in short PPM instances as the adding series is 0 for the
first timer after about 110 segments.

On first sight, C3 (cf. \figurename~\ref{fig:cluster3}) seems to be between
C1 and C2. The adding curve is mostly situated between 0.4 and 0.7, a littler
lower than for C2, but still higher compared to C1. Similar values can be
observed for the reconciliation curve. The deleting curve remains below 0.1.
The duration of the PPM instances is also between the duration of C1 and C2
as the adding series is 0 for the first time after about 137 segments.

Table~\ref{tab:statistics} presents general statistics on the number of
adding operations, the number of deleting operations, the number of
reconciliation operations and the total number of operations for each
cluster. Interestingly, modelers in C1 had more adding operations, more
deleting operations and, probably most notable, almost twice as many
reconciliation operations compared to C2 and C3. At a first glance, the
numbers for C2 and C3 appear to be very similar.

The following procedure for conducting the statistical analysis was used.  If the
data was normally distributed and homogenity of variances was given we used
Oneway ANOVA to test for differences among the groups. Pairwise comparisons were
done using the bonferroni post-hoc test. Note that the bonferroni post-hoc test
uses an adapted significance level. Therefore, p-values less than $0.05$ are
considered to be significant, i.e., there is no need to divide the significance
level by the number of groups, i.e., clusters. In case normal distribution or
homogenity of variance was not given a non-parametric alternative to ANOVA, i.e.,
kruskall-wallis, was utilized to test for differences among the groups. Pairwise
comparisons were done using the t-test for (un)equal variances (depending on the
data) if normal distribution was given. If no normal distribution could be
identified the mann-whitney test was utilized. In either case, i.e., t-test or
mann-whitney test, the bonferroni correction was applied, i.e., the significance
level was divided by the number of clusters.

The results are summarized in Table~\ref{tab:statisticsSummary}, indicating
significant differences between C1 and C2 and C1 and C3, but not between C2 and
C3. Only significant differences are stated.

\begin{table}[t]\sf\small
\begin{threeparttable}
\begin{centering}
\begin{tabular}{@{\extracolsep{.5em}}lp{0.7cm}lp{1.2cm}p{1.2cm}p{1.2cm}}
\hline
\textbf{Statistic} & & \textbf{All groups} & \multicolumn{3}{l}{\textbf{Pairwise comparison}} \\
\cline{4-6}
& & & 1-2 & 1-3 & 2-3\\
\hline
Number of Adding Operations & Sig. & 0.000\tnote{a} & 0.003\tnote{a} &
0.000\tnote{a} & \\
\cline{2-6}
& test &  Oneway ANOVA & \multicolumn{3}{l}{Bonferroni post-hoc test}
\\
\hline
Number of Deleting Operations & Sig. & 0.000\tnote{a} & 0.000\tnote{b} &
0.000\tnote{b} & \\
\cline{2-6}
& test & Kruskall-wallis & \multicolumn{3}{l}{Mann-whitney test} \\
\hline

Number of Reconciliation & Sig. & 0.000\tnote{a} & 0.000\tnote{b} &
0.000\tnote{b} &\\
\cline{2-6} Operations & test & Kruskall-wallis & \multicolumn{3}{l}{t-test
for unequal
 variances}
\\
\hline

Number of Total Operations & Sig. & 0.000\tnote{a} & 0.000\tnote{b} &
0.000\tnote{b} &
\\
\cline{2-6}
& test & Kruskall-wallis & \multicolumn{3}{l}{t-test for unequal variances} \\
\hline
\end{tabular}
\begin{tablenotes}[para]
\item[a]{$p < 0.05$}
\item[b]{$p < 0.05/3$}
\end{tablenotes}
\caption{Significant differences for statistics}
\label{tab:statisticsSummary}
\end{centering}
\end{threeparttable}
\end{table}

\subsection{Applying Measures}

In order to further distill the properties of the three clusters, we
calculated the measures described in Section~\ref{sec:measures} for each
PPM. Table~\ref{tab:ppm_measures} provides an
overview presenting the average values for each measure in each cluster. As
indicated in \figurename~\ref{fig:cluster1}, C1 constitutes the highest
number of PPM iterations. Tightly connected to this observation is the
average iteration chunk size. Modelers in C2 added by far the most
content per iteration to the process model. Also the number of iterations
containing delete iterations is higher for C1 than for the other clusters,
which is consistent with the higher number of delete operations (cf.
Table~\ref{tab:statistics}). The amount of time spent on comprehending the
task description and developing the plan on how to incorporate them into the
process model seems to be far larger for C1 compared to C2, which has the
lowest share of comprehension, but also larger compared to C3. When
considering reconciliation breaks C3 sets itself apart posting the lowest
number of reconciliation breaks. C2 is somewhere in between and C1 has the
highest number of reconciliation breaks.

\begin{table}[tb]\sf\small
\centering
\begin{tabular}{@{\extracolsep{.5em}}lrrr}
\hline
\textbf{Measure} & \textbf{C1} & \textbf{C2} & \textbf{C3} \\
\hline
Avg. no. of PPM iterations & 21.50 & 12.32 & 14.69\\
Avg. iteration Chunk Size & 3.66 & 5.28 & 4.24 \\
Avg. share of comprehension & 49.88 & 39.28 & 45.02 \\
Avg. reconciliation breaks & 21.37 & 18.14 & 13.85 \\
Avg. delete iterations & 17.06 & 10.07 & 10.83 \\
\hline
\end{tabular}
\caption{Measures per cluster}
\label{tab:ppm_measures}
\end{table}

Statistical analysis of the differences between the groups was performed
following the procedure described in the previous section. An overview of the
results is presented in Table~\ref{tab:measuresSummary}. Only significant
differences are stated. In constrast to the statistics presented in
\tablename~\ref{tab:statisticsSummary}, we were able to identify significant
differences between C2 and C3.

\begin{table}[t]\sf\small\centering
\begin{threeparttable}
\begin{tabular}{@{\extracolsep{.5em}}lp{0.7cm}lp{1.2cm}p{1.2cm}p{1.2cm}}
\hline
\textbf{Measure} & & \textbf{All groups} & \multicolumn{3}{l}{\textbf{Pairwise
comparison}} \\
\cline{4-6}
& & & 1-2 & 1-3 & 2-3\\
\hline
Iteration Chunk Size & Sig. & 0.000\tnote{a} & 0.000\tnote{b} &
0.000\tnote{b} & 0.007\tnote{b} \\
\cline{2-6}
& test &  Kruskall-wallis & \multicolumn{3}{l}{t-test for unequal variances}
\\
\hline
Number of Iterations & Sig. & 0.000\tnote{a} & 0.000\tnote{b} & 0.000\tnote{b} &
0.004\tnote{b} \\
\cline{2-6}
& test & Kruskall-wallis & \multicolumn{3}{l}{t-test for unequal variances} \\
\hline
Share of Comprehension & Sig. & 0.000\tnote{a} & 0.000\tnote{a} & 0.036\tnote{a}
& 0.045\tnote{a}
\\
\cline{2-6}
& test & Oneway ANOVA & \multicolumn{3}{l}{Bonferroni post-hoc test} \\
\hline
Delete Iterations & Sig. & 0.005\tnote{a} & 0.026\tnote{a} & 0.011\tnote{a} & \\
\cline{2-6}
& test & Oneway ANOVA & \multicolumn{3}{l}{Bonferroni post-hoc test} \\
\hline
Reconciliation Breaks & Sig. & 0.005\tnote{a} & & 0.004\tnote{a} & \\
\cline{2-6}
& test & Oneway ANOVA & \multicolumn{3}{l}{Bonferroni post-hoc test} \\
\hline
\end{tabular}
\begin{tablenotes}[para]
\item[a]{$p < 0.05$}
\item[b]{$p < 0.05/3$}
\end{tablenotes}
\caption{Significant differences for measures}
\label{tab:measuresSummary}
\end{threeparttable}
\end{table}

\vspace{-0.3cm}
\section{Discussion}\label{sec:interpretation}
In this section we present our insights when comparing the identified clusters
and we discuss the lessons learned in this work and how they influence our
future work. Additionally, limitations of this work are described.

\subsection{Cluster C1}

C1 can be clearly distinguished from C2 and C3. This becomes evident on
visual inspection of \figurename~\ref{fig:cluster1}, but also when
considering the number of adding operations, the number of deleting
operations, the number of reconciliation operations and the total number of
operations. We identified statistically significant differences between C1
and C2 and between C1 and C3 for all statistics (cf.
Table~\ref{tab:statisticsSummary}).

In general, modelers in C1 had rather long PPM instances, i.e., the number of PPM
iterations was significantly higher compared to C2 and C3. In addition, modelers
in C1 spent more time on comprehension compared to C2. Modelers started rather
slowly, not eclipsing 0.5 adding operations. The slow modeling speed is
underlined by the significantly lower chunk size compared to C2 and C3. During
the whole process, adding operations are accompanied by a relatively high amount
of delete operations. This is underlined by the significant differences in the
number of delete iterations between C1 and C2 and C1 and C3 (cf.
Table~\ref{tab:measuresSummary}). Also, we observed a fairly large amount of
reconciliation operations, culminating in a massive peak after about half of the
PPM instances.

The results suggest that modelers in C1 were not as goal oriented as their
colleagues in other clusters, since they spent a great amount of time on
comprehension, added more modeling elements which were subsequently removed
and put significantly more effort into improving the visual appearance of the
process model. There might be multiple reasons for this behavior. On the one
hand, it could point toward modelers having trouble executing the modeling
task and therefore needed more reconciliation to facilitate their
understanding of the process model at hand. On the other hand, their focus on
layouting might have acted as a distraction from the modeling task, resulting
in the higher number of adding operations and deleting operations. Still,
other techniques will be required for further investigating this claim, e.g.,
think aloud protocols (cf.~\cite{ErSi93}).

\subsection{Cluster C2}
When inspecting \figurename~\ref{fig:cluster2} the very steep start of the
adding curve strikes the eye, indicating that modelers started creating the
process model right away. When focusing on reconciliation operations, several
spikes in the layouting curve can be identified, notably one last spike right
after the number of adding operations decreases. As already mentioned above,
C2 is statistically significant different compared to C1 for all statistics
presented in Table~\ref{tab:statistics}. No differences can be identified
between C2 and C3.

Considering the measures described in Section~\ref{sec:measures}, C2 has a
significantly higher chunk size compared to C1 and C3. Similarly, we observed
the lowest number of PPM iterations. This means that modelers add a lot more
content per PPM iteration. In addition, modelers
in C2 did not spend as much time on comprehension compared to
modelers in C1 and C3.

In a nutshell, modelers of C2 are very focused and goal oriented following a
straight path when creating the process model. They are quick in making
decisions about how to proceed and only slow down their modeling endeavor
from time to time for some reconciliation, resulting in short PPM
instances.

\subsection{Cluster C3}
\figurename~\ref{fig:cluster3} shows the PPM instances for C3. The processes
are shorter compared to C1 and longer compared to C2. It is
lacking the fast start of the adding curve we identified for C2. The
reconciliation curve is more or less following the adding curve. Notably,
this is the only curve without a reconciliation spike once the number of
adding operations decreases.

The calculated measures indicate clear differences to C1 when it comes to
chunk size, number of iterations, share of comprehension, but also
number of delete operations and reconciliation breaks. C2 and C3 differ
in chunk size, the number of iterations and the time spent on comprehension.

\begin{wrapfigure}{r}{0.5\textwidth}
\vspace{-15pt}
\begin{center}
  \includegraphics[width=0.5\textwidth]{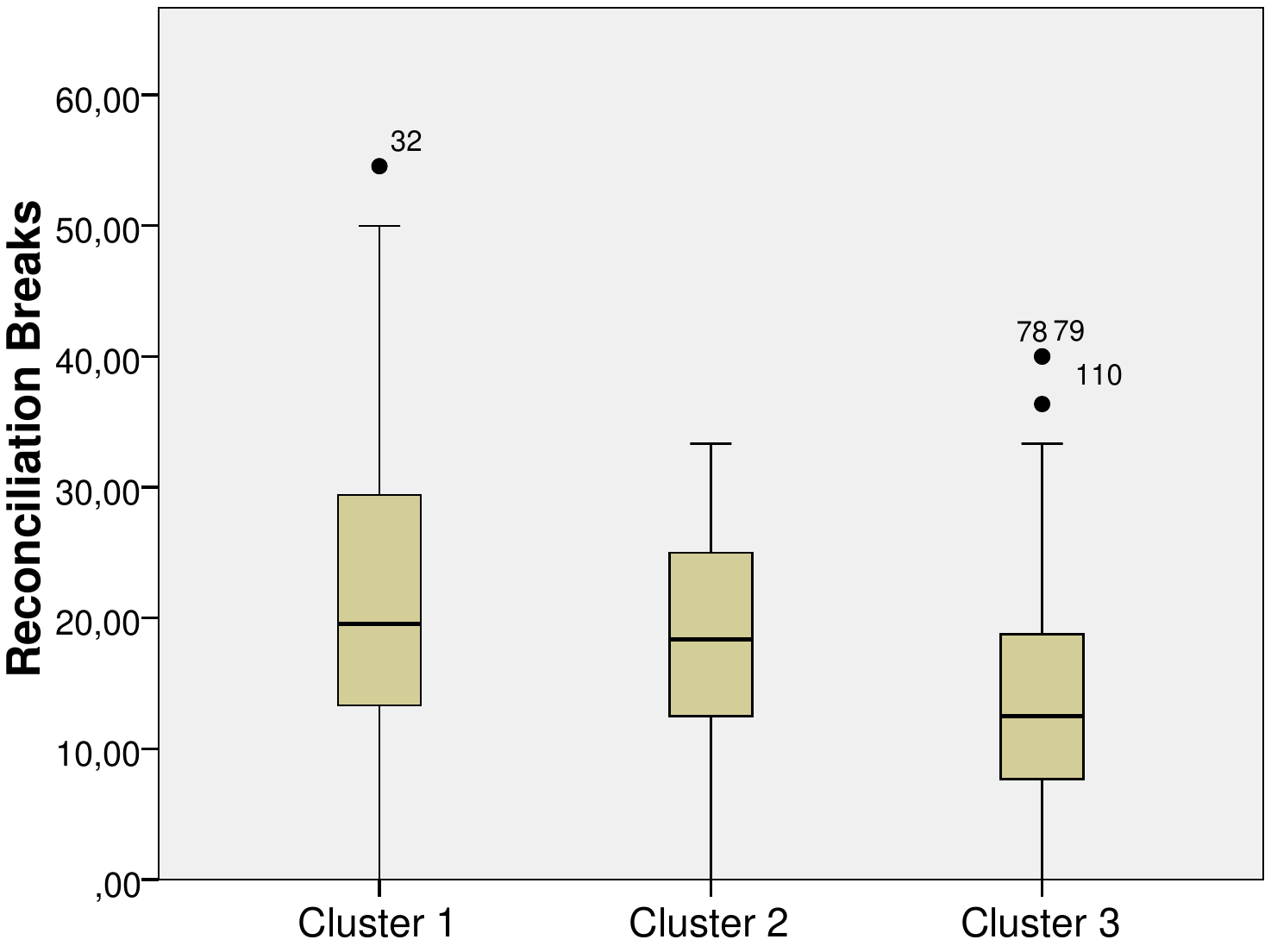}
  \caption{Reconciliation Breaks}
  \label{fig:reconiliationBreaks}
\end{center}
\vspace{-15pt}
\end{wrapfigure}

When comparing C2 and C3, the question arises whether modelers in C3 followed
the same strategy as modelers in C2, just a little slower. We believe that,
in contrary to C2, modelers of C3 followed a more systematic approach to
process modeling. They continuously reconciled their process model,
alleviating the need for dedicated reconciliation breaks. This is indicated
by the lack of a reconciliation spike after the decrease of adding operations
in \figurename~\ref{fig:cluster3}. Additionally, reconciliation breaks
points into this direction (18.16 for C2 vs. 13.85 for C3).
\figurename~\ref{fig:reconiliationBreaks} depicts the reconciliation breaks
box plot, hinting at a difference in reconciliation breaks between C2 and C3.
Still, the difference did not turn out to be statistically significant
leaving us with some future work on investigating whether this claim actually
holds.

Additionally, we believe that different reasons for reconciliation breaks exist.
On the one hand, modelers are forced to stop their modeling endeavour and layout the
process model when they are overwhelmed by the complexity at hand. On the other
hand, some modelers might stop modeling at strategic points to reconcile the
process model in order to avoid situations like the one mentioned above before they even
arise. Even though this explanation would fit the boxplot depicted in
\figurename~\ref{fig:reconiliationBreaks}, further investigations are in demand
to fully understand the reconciliation behavior of modelers.

\subsection{Lessons Learned}
We were able to identify three different modeling styles using cluster analysis.
Differences among the clusters were subsequently validated using a series of
measures quantifying the PPM. Note that these measures
were defined prior to performing the cluster analysis. The measures are based on
the detected iterations of the PPM, approaching the
PPM from a different angle. Therefore they enable us to
validate the differences among the three clusters.

The detected modeling styles contribute to our understanding of the process of
process modeling, as, to our knowledge, this is the first systematic attempt to
establish a categorization of PPM instances in the domain of
business process modeling. We believe that further refinements of the
categorization will emerge, ultimately enabling us to create personalized
modeling environments based on their observed modeling behavior. In addition,
these findings can be exploited for teaching purposes. For example, teachers
might be able to identify students facing difficulties during a modeling
assignment based on their modeling behavior and provide them with additional
support. Still, some research questions emerging from these findings have to be
addressed first. The most pressing might be whether a modeler's personal style
persists over several different modeling tasks or if the modeling style is
determined by the modeling task at hand. To answer this question further
empirical investigations are in demand. Based on some preliminary observations
we would assume that the influence of the modeling task cannot be
neglected. Even though a modeler might like to create a process model in a
straight forward, goal oriented way, the complexity of the modeling task might
force her to reduce the modeling speed and switch to a more conservative modeling
style.

On a long-term basis questions on how to exploit this knowledge to improve
the quality of the resulting process model become evident. Unfortunately, the
naive assumption that one modeling style is superior to the others could not
be confirmed. All clusters contained excellent process models and process
models of low quality. This is not surprising though. Even modelers in C1 who
face difficulties, exhibiting long PPM instances, can still come up with good
process models if they succeed in overcoming the adversity they are facing.

\vspace{-0.3cm}

\subsection{Limitations}
The interpretation of our findings is presented with the explicit acknowledgement
of a number of limitations to our study. First of all, our respondents
represented a rather homogeneous and inexperienced group. Although relative
differences in experience were notable, the group is not representative for the
modeling community at large. At this stage, in particular, the question can be
raised whether experienced modelers also exhibit the same style elements as
skillful yet inexperienced modelers. In other words, will experienced modelers
display similar characteristics of style or can other styles be observed within
their approaches? Note that we are mildly optimistic about the usefulness of the
presented insights on the basis of modeling behavior of graduate students, since
we have established in previous work that such subjects perform equally well in
process modeling tasks as some professional modelers~\cite{ieee10-reijers}.

Secondly, the influence of the modeling task---more precisely, the modeling
task's complexity (cf.~\cite{Card08})---on the PPM is not fully understood. All
students in our modeling session were working on the same modeling assignment.
Hence, the observed clusters might be specific to modeling tasks of this
complexity level. Further investigations will be necessary to let sunlight fall
on the influence of the modeling task, which might result in the emergence of
additional clusters. Preliminary results of a different modeling task suggest the
existence of modeling styles comparable to the results presented in this paper.

Thirdly, we can not rule out that KMeans identified a local
minimum, resulting in a suboptimal clustering. To counter this threat we
validated the clustering using a series of measures quantifying the PPM and
identified significant differences among the three groups.

\vspace{-0.2cm}
\section{Related Work}\label{sec:relatedWork}

Our work is essentially related to model quality frameworks and research on the
process of modeling.

There are different frameworks and guidelines available that define quality for
process models. Among others, the SEQUAL framework uses semiotic theory for
identifying various aspects of process model quality~\cite{krogstie06}, the
Guidelines of Process Modeling describe quality considerations for process models
\cite{GOPM}, and the Seven Process Modeling Guidelines define desirable
characteristics of a process model \cite{DBLP:journals/infsof/MendlingRA10}.
While each of these frameworks has been validated empirically, they rather take a
static view by focusing on the resulting process model, but not on the act of
modeling itself. Our research takes another approach by investigating the
process followed to create the process model.

Research on the process of modeling typically focuses on the interaction between
different parties. In a classical setting, a system analyst directs a domain
expert through a structured discussion subdivided into the stages elicitation,
modeling, verification, and validation \cite{hoppenbrouwers05er,frederiks06}. The
procedure of developing process models in a team is analyzed in 
\cite{rittgen07-caise} and characterized as a negotiation process. Interpretation
tasks and classification tasks are identified on the semantic level of modeling.
Participative modeling is discussed in \cite{stirna07-caise}. These works build
on the observation of modeling practice and distill normative procedures for
steering the process of modeling towards a good completion. Our work, in turn, 
focuses on the formalization of the process model, i.e., the modeler's interactions with
the modeling environment when creating the formal business process model. 

\vspace{-0.5cm}
\section{Summary}\label{sec:summary}
This paper contributes to our understanding of the PPM as
it constitutes the first systematic attempt to identify different modeling styles
in the domain of business process modeling. We conducted a modeling session with
115 students of courses on business process management, collecting their
PPM instances. We were able to identify three different modeling
styles using cluster analysis and validated the retrieved clusters using a series
of measures for quantifying the PPM. We believe that a
better understanding regarding the PPM will be beneficial for future process
modeling environments and will support teachers in mentoring their students on
their way to professional process modelers.

\footnotesize 
\noindent\textbf{Acknowledgements.} This research was funded by the Austrian
Science Fund (FWF): P23699-N23. The research stay of Dr. Pnina Soffer in
Innsbruck, Austria was funded by BIT School.

\bibliographystyle{splncs}
\bibliography{literature}

\end{document}